\documentclass[prb,twocolumn,superscriptaddress,showpacs]{revtex4}
\usepackage{amssymb}
\usepackage{graphicx}
\usepackage[dvips]{color}
\definecolor{MyOrange}{rgb}{1,0.6471,0}
\begin{document}
\title{Equation of state and Raman-active $E_{2g}$ lattice phonon in phases I, II, and III  of solid hydrogen and deuterium}
\author{Yu. A. Freiman}
\email{freiman@ilt.kharkov.ua} \affiliation{B.Verkin Institute for
Low Temperature Physics and Engineering of the National Academy of
Sciences of Ukraine, 47 Lenin avenue, Kharkov, 61103, Ukraine}
\author{Alexei Grechnev}
\affiliation{B.Verkin Institute for Low Temperature Physics and
Engineering of the National Academy of Sciences of Ukraine, 47
Lenin avenue, Kharkov, 61103, Ukraine}
\author{S. M. Tretyak}
\affiliation{B.Verkin Institute for Low Temperature Physics and
Engineering of the National Academy of Sciences of Ukraine, 47
Lenin avenue, Kharkov, 61103, Ukraine}
\author{Alexander F. Goncharov}
\affiliation{Geophysical Laboratory, Carnegie Institution of
Washington, 5251 Broad Branch Road NW, Washington, DC 20015, USA}
\author{Russell J. Hemley}
\affiliation{Geophysical Laboratory, Carnegie Institution of
Washington, 5251 Broad Branch Road NW, Washington, DC 20015, USA}

\date{\today}

\begin{abstract}
{We present results of lattice dynamics calculations of the $P-V$
equation of state and the pressure dependence of the Raman-active
$E_{2g}$ lattice phonon for  $p-$H$_2$ and $o-$D$_2$ in  a wide
pressure range up to $\sim$2 Mbar using our recently developed
semi-empirical many-body potential, and density-functional theory.
Comparison with existing body of experimental and theoretical
results showed that the employed many-body potential is a reliable
basis for high-precision calculations for phases I, II, and III of
solid hydrogens.}

\pacs{64.30.Jk, 67.80.F-, 78.30.Am}
\end{abstract}
\maketitle
\begin{large}
An accurate determination of the equation of state (EOS) of solid
hydrogens has been an important research objective for decades.
Systematic high-pressure studies were started in the seventies of
the last century \cite{And74,Sil78a,Dri79} (see reviews
\cite{Sil80,Mao94,Man97} and references therein). At present these
x-ray and neutron studies span the pressure range up to $\sim$2
Mbar \cite{Ish83,Ish85,Haz87,vS88,Mao88,Hem90,Bes90,Lou96,Aka10}
and temperature range up to 1000 K. The highest compression
reached in the EOS experiments is 10.4 for solid H$_2$
\cite{Aka10} (7.6 for solid D$_2$ \cite{Lou96}), essentially
higher than that for solid helium (8.4) \cite{Frei09}. The EOS
data provide a fundamental basis for examining intermolecular
interactions, and for testing {\it ab initio} theories. A number
of model intermolecular potentials  \cite{Sil78,Ross83} have been
proposed based on the experimental EOS data. Another experimental
technique which complements x-ray and neutron diffraction by
providing direct information on intermolecular interactions and
vibrational dynamics is Raman scattering. The hcp structure has a
Raman-active optical mode ($E_{2g}$ symmetry) in the phonon
spectrum which corresponds to the out-of-phase shear motions in
the two orthogonal directions in the $ab$ plane. The frequency
range of this Raman mode is extremely large, from 36 cm$^{-1}$ at
zero pressure
\cite{Sil72,Niel73,Berk79,Wijn83,Lag85,Hem90a,Han94,Gon98,Gon01}
to ~ 1100 cm$^{-1}$ at ~250 GPa. The Raman spectrum  of solid
molecular deuterium has been measured up to $\sim$200 GPa
\cite{Sil72,Niel73,Berk79,Wijn80,Wijn83,Lag85,Hem93}. These
measurements show that hcp-based structures are stable in this
pressure range. The calculations of the $E_{2g}$ Raman frequency
$\nu(P)$ using various empirical potentials
\cite{Wijn83,Lag85,Mao94} show that the result is highly sensitive
to details of the potential used. Therefore, comparing the
calculated and experimental $\nu(P)$ is a hard test for any
empirical potential (or for any other theoretical method, e.g.
\emph{ab initio} calculations). It is essential that these
properties are sensitive to different characteristics of the
intermolecular potential: while EOS is sensitive to the potential
well depth, the Raman scattering experiment probes the second
derivative  of the potential at the minimum. In our recent paper
\cite{Frei11} we have proposed new semi-empirical isotropic
potentials for H$_2$ and D$_2$. Unlike the previous potentials
\cite{Sil78,Ross83,Hem90,Duf94} they include not only pair forces,
but triple forces as well. The goal of the present paper is to
perform detailed calculations of the EOS and pressure dependence
of Raman frequencies for H$_2$ and D$_2$ using our new potentials
and to compare the results to available  experimental data and
theoretical results for a wide pressure range which spans the
phases I, II, and III of solid $p$-H$_2$ and $o$-D$_2$.

As mentioned above, our potential includes pair (U$_p$) and triple
($U_{\rm tr}$) intermolecular forces ($U_{\rm tot} = U_p + U_{\rm
tr}$).  This potential was designed in a manner similar to the
potential for solid helium \cite{Lou87,Frei08}. It has the form of
a sum of the pair Silvera-Goldman (SG) potential \cite{Sil78}
(discarding the $R^{-9}$ term) and three-body terms which include
the long-range Axilrod-Teller dispersive interaction and a
short-range three-body exchange interaction. The latter was used
in a Slater-Kirkwood form \cite{Lou87,Frei08}. Our potential also
includes the translational-rotational interaction however we have
found that its contributions both to EOS and Raman frequencies are
negligible. An explicit form and parameters of the employed
potential are given in Ref. \cite{Frei11}. We restrict ourselves
to $T = 0$ K, with the zero-point energy taken into account using
the Einstein approximation. A small pressure range ($\sim 0.5$
GPa) where quantum-crystal effects play a decisive role was
excluded from consideration.

The decomposition of the total energy into contributions from the
pair forces (E$_p$), triple forces (E$_{\rm tr}$),
 and the zero-point energy (E$_{zp}$) is presented in Fig. 1.
  The inset of Fig. 1 shows the respective decomposition
 of EOS. As can be seen, the interplay between these
 three contributions is rather complicated and
different for different regions of the molar volume. At relatively
small compressions $V_0/V < 2$, the ground-state energy (and
consequently the EOS) is dominated by the zero-point energy
$E_{zp}$. The zero-point contribution decreases with rising
compression, but it remains significant up to the highest
pressures reached in EOS experiments. At the ten-fold compression
the relative contribution $E_{zp}/E_{gs}$ remains as high as 20\%.
It is interesting to note that for the eight-fold compression
$E_{zp}/E_{gs}$ coincides with that for solid $^4$He, but for
helium it increases with decreasing compression more steeply and
already at three-fold compression the zero-point contribution
dominates in the ground-state energy \cite{Frei09}. The three-body
attraction becomes appreciable at the compressions higher than
two-fold which corresponds to pressures about 2 GPa. The relative
contribution of the tree-body forces $|E_{{\rm tr }}|/E_p$
monotonically increases with pressure and for the eight-fold
compression it reaches 0.5.

\begin{figure}
\includegraphics[scale=0.72]{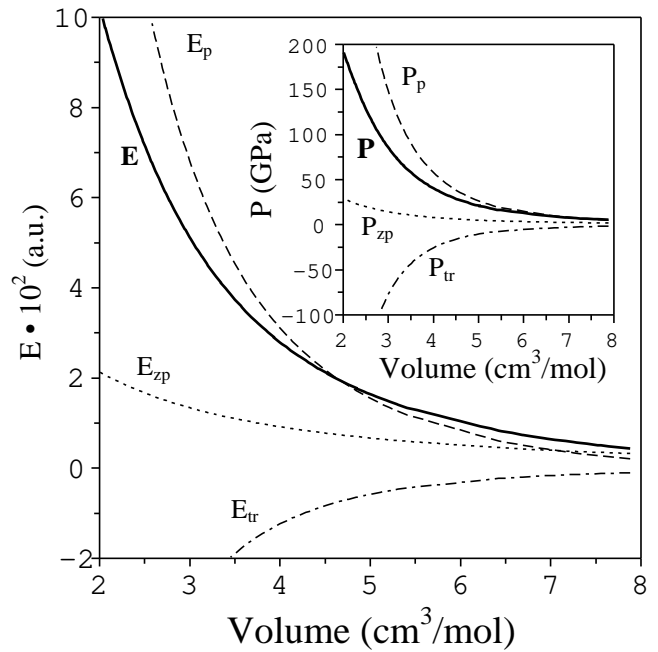}
\caption{Contributions of the pair $E_p$ and triple $E_{\rm tr}$
forces, and zero-point energy $E_{zp}$ to the total ground-state
energy $E_{gs}$ for solid $p$-H$_2$. The inset shows the
respective contributions to EOS.}
\end{figure}

There has been many attempts to propose effective pair potentials
which would have the same softening effect as attractive many-body
forces \cite{Ross83,Hem90,Duf94}. To account for these effects
Hemley {\it et al} modified the SG potential \cite{Sil78} with a
short-range correcting term \cite{Hem90,Duf94}. This
Hemley-Silvera-Goldman effective potential was shown to fit static
compression data up to 40 GPa. The $P(V)$ calculated with this
effective pair potentials for higher pressures \cite{Lou96}
increases far more rapidly than in experiment.

\begin{figure}
\includegraphics[scale=0.72]{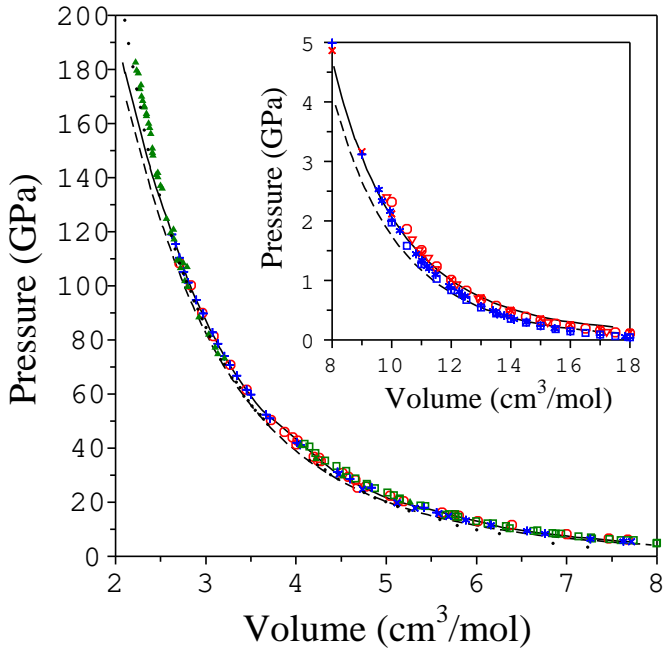}
\caption{(Color online) Calculated and experimental
pressure-volume relations for solid H$_2$ and D$_2$.
Semi-empirical calculations for many-body potential (this work):
$p-$H$_2$ (solid line), $o-$D$_2$ (dashed line); DFT-GGA
calculations ($\cdot \cdot \cdot$) \cite{Zhan07}. Experiment
(reduced to 0 K): (H$_2$:
{\color{green}$\blacktriangle$}\cite{Aka10}, {\color{red} $\circ$}
\cite{Lou96}, {\color{green} $\square$} \cite{Mao94}); (D$_2$:
{\color{blue} $+$} \cite{Lou96}, {\color{blue} $\times$}
\cite{Bes90}, {\color{blue} $\ast$} \cite{Mao94}; the inset shows
the small-pressure range. Experiment (reduced to 0 K): (H$_2$:
{\color{red} $\circ$} \cite{Dri79}, {\color{red} $\times$}
\cite{vS88}, {\color{red} $\triangledown$} \cite{Ish83}); (D$_2$:
{\color{blue} $\square$} \cite{Dri79}, {\color{blue} $+$}
\cite{vS88}, {\color{blue} $\ast$} \cite{Ish85}.}
\end{figure}

The calculated equations of state $P(V)$ for solid hydrogen and
deuterium  are shown in Fig. 2 in comparison with DFT-GGA
calculations \cite{Zhan07} and the experimental results from Refs.
\cite{Dri79,Ish83,Ish85,vS88,Lou96,Hem90,Mao94,Duf94,Aka10}. As
can be seen, the semi-empirical calculation with the proposed
many-body potential is in an excellent agreement with  experiment
in the pressure range 1 - 140 GPa (phases  I and II). These
results can be compared favorably with recently published EOS
calculations \cite{Cail11}. From 140 GPa onwards, the theoretical
$P(V)$ curve lies slightly  below the experimental one, and at the
maximum pressure of 180 GPa (Phase III) the difference grows up to
about $10\%$. The reason for this is the neglect of the higher
order ($n>3$) terms in the $n$-body expansion. The effect of the
large-$n$ terms increases with pressure, and at the metallization
point the $n$-body expansion would converge extremely slowly.
Methods based on the density functional theory (DFT) within local
density approximation (LDA) and generalized  gradient
approximation (GGA) are somewhat of an opposite to the empirical
potentials method. Indeed, the accuracy of the EOS from DFT-GGA
\cite{Zhan07} improves with the increase of pressure: in the
pressure range 180-70 GPa the EOS from GGA practically coincides
with the experimental one and for $P>140$ GPa the agreement is
better than for our empirical potentials, but at lower pressures
the {\it ab initio} results progressively underestimate the
pressure, and GGA gives a strongly underestimated equilibrium
volume of about 8 cm$^3$/mol. The reason for this is twofold:
first, GGA gives a poor description of the van der Waals forces,
and second, DFT calculations ignore all quantum zero-point motions
of nuclei, including the distinction between ortho- and
para-species. We also compare theoretical and the experimental
values for the isotopic shift $\Delta P(V) \equiv P_{\rm
D_2}(V)-P_{\rm H_2}(V)$. In accordance with Ref. \cite{Lou96} we
find that the empirical potentials strongly overestimate $\Delta
P$.

\begin{figure}
\includegraphics[scale=0.31]{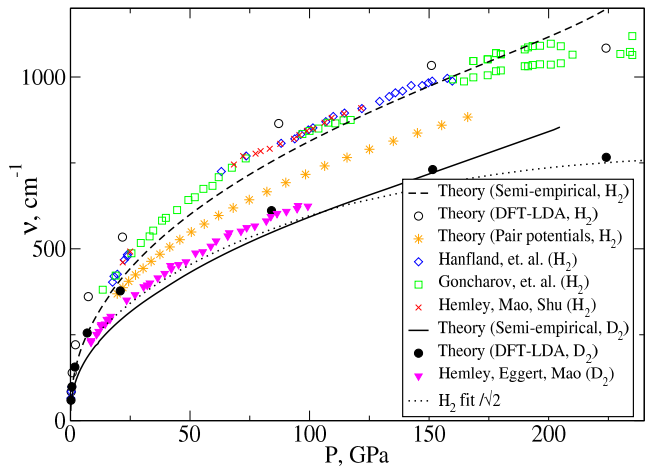}
\caption{(Color online) Calculated and experimental Raman
frequencies as a function of pressure for solid hydrogen and
deuterium.  Semi-empirical calculations for many-body potential:
this work. $p-$H$_2$ (dashed line),  calculations for the SG
potential {\color{MyOrange} $\ast$} \cite{Mao94}; $o-$D$_2$ (solid
line). DFT-LDA theory (this work): H$_2$ $\circ$, D$_2$ $\bullet$.
Experiment (H$_2$): {\color{blue} $\diamond$} \cite{Han94},
{\color{red} $\times$} \cite{Hem90a}), {\color{green} $\square$}
\cite{Gon98,Gon01}; (D$_2$): {\color{red} $\triangledown$}
\cite{Hem93}.}
\end{figure}

The comparison between theoretical and experimental pressure
dependencies $\nu(P)$ of the $E_{2g}$ optical phonon Raman-active
mode is presented in Fig. 3. Since we could not find any DFT data
of this mode in the literature, we have also calculated the
$E_{2g}$ Raman frequency of solid H$_2$ and D$_2$ in the Pca2$_1$
structure - one of the plausible candidates for the orientational
structure of phases II and III - using DFT-LDA approximation. Our
calculations were done using the Full-Potential Linear Muffin-Tin
Orbital (FP-LMTO) code RSPt \cite{wills-book}.

Comparing the theoretical results with the experiment we see that
similarly to that we had for the EOS at pressures lower than
$\sim$150 GPa the semi-empirical curves agree with experiment
better than DFT calculations but at higher pressures the situation
is reversed. The limiting pressures at which the semi-empirical
approach still works are $\sim$175 GPa while LDA has a fine
agreement with the experiment for H$_2$ and with frequencies
obtained for D$_2$ with the help of harmonic ratio of $\sqrt2$
from 150 GPa up to the highest considered pressures $\sim$230 GPa.
The frequencies calculated from the SG potential \cite{Sil78}
deviate from experiment even for very low pressures. The same is
true \cite{Mao94} for the effective HSG pair potential
\cite{Hem90,Duf94}. Thus we have shown that while effective pair
potentials work reasonably well for EOS up to 40 GPa, they fail
for the dynamical properties like Raman spectrum, where the
explicit inclusion of the 3-body forces is necessary.

In conclusion, we have calculated the EOS and the pressure
dependence of the Raman-active $E_{2g}$ mode using our recently
proposed many-body potentials \cite{Frei11}, and compared the
results to the experiment and previous semi-empirical and DFT
calculations. Also, DFT-LDA calculations of the $E_{2g}$ Raman
frequency were performed. For phases I and II ($P<150$ GPa) the
proposed many-body potentials give excellent agreement with the
experiment, much better then any previous calculations. It proves
that the new potentials are a reliable basis for high-precision
calculations of structure and dynamics of H$_2$ and D$_2$ up to
about 140 GPa. In particular, they provide a huge improvement over
any effective two-body potentials, stressing the importance of
including the 3-body forces. For the higher pressures (Phase III)
the DFT approach is preferable.

\begin{flushleft}

\end{flushleft}
\end{large}
\end{document}